# Molecular Beam Epitaxy of Twin-Free $Bi_2Se_3$ and $Sb_2Te_3$ on $In_2Se_3$/InP(111)B Virtual Substrates

*Kaushini S. Wickramasinghe[1], Candice Forrester[1,2], Maria C. Tamargo[1,2] ***

[1]Department of Chemistry, The City College of New York, NY, NY 10031

[2]Chemistry Program, CUNY Graduate Center, NY, NY 10016

*Electronic mail: mtamargo@ccny.cuny.edu

Three-dimensional topological insulators (3D-TIs) are a new generation of materials with insulating bulk and exotic metallic surface states that facilitate a wide variety of ground-breaking applications. However, utilization of the surface channels is often hampered by the presence of crystal defects, such as antisites, vacancies and twin domains. For terahertz device applications, twinning is shown to be highly deleterious. Previous attempts to reduce twins using technologically important InP(111) substrates have been promising, but have failed to completely suppress twin domains while preserving high structural quality. Here we report growth of twin-free molecular beam epitaxial $Bi_2Se_3$ and $Sb_2Te_3$ structures on ultra-thin $In_2Se_3$ layers formed by a novel selenium passivation technique during the oxide desorption of smooth, non-vicinal InP(111)B substrates, without the use of an indium source. The formation of un-twinned $In_2Se_3$ provides a favorable template to fully suppress twin domains in 3D-TIs, greatly broadening novel device applications in the terahertz regime.

Three dimensional topological insulators (3D-TIs) have attracted a great deal of interest in the past decade due to their non-trivial topology, which gives rise to metallic surface states protected by time reversal symmetry, and an insulating bulk.[1,2,3,4] A wide variety of applications in thermoelectrics[5], spintronics[6], twistronics[7,8], and quantum computation[9,10] are being considered. They also provide a fundamental platform to explore novel physics.[11] The topological nature of the surface states also enables a range of other novel technologies such as polarization selective terahertz detectors[12,13,14]. In these, the single crystal nature of the TIs becomes of paramount importance.

Among the 3D TIs, $Bi_2Se_3$ and $Sb_2Te_3$ are most actively pursued due to their experimentally verified single Dirac cone at the Γ point.[15] Their crystal structure is rhombohedral with a space group $D_{3d}^5 (R\bar{3}m)$, with five atoms per unit cell.[2] They are layered materials that exhibit a quintuple layer (QL) structure consisting of Se-Bi-Se-Bi-Se atoms along the c-direction. The QLs are separated by a van der Waals gap. Because of their van der Waals nature, there is a wide range of substrates that can be used for the epitaxial growth of these TIs.[16,17,18]

However, the influence of the substrate on the molecular beam epitaxy (MBE) grown material properties has been shown to be non-negligible.[19,20] It has been shown that in some cases there is a chemical interaction of the atoms between the substrate and the material at the interface which gives rise to quasi van der Waals growth,[19,20] or other times an amorphous disordered interfacial layer is present[21]. Further, there is a large unintentional background doping that conceals their surface channels[21,16] mainly due to the above mentioned interface issue, as well as to the inherent low-energy native defects[22] (Se or Te vacancies and antisite defects). Thus the device and transport studies of these materials have been hindered.[16,23] Many efforts to reduce background doping have been reported; for example improving the interface via different substrate preparation techniques,[24,25] and growing a buffer layer of $In_2Se_3$[25] or $In_xBi_{1-x}Se_3$[26,27] has shown a significant reduction in defects and background doping in $Bi_2Se_3$.

Moreover, crystal defects such as twin domains[28,29] are frequently observed regardless of the substrate or the buffer layer that is being used. In some applications, twinning in these layers is shown to be more deleterious to topological properties than intrinsic bulk doping. For example, twinning reduces helicity dependent topological photocurrent, thus, eliminating twinning can provide a path to chip-scale polarimeters[14], among other devices.

Most often, the 3D TIs are grown on sapphire substrates due to sapphire's low cost and inert chemical nature. InP(111), a substrate of considerable technological importance, and its surface provide the appropriate symmetry and lattice constant for $Bi_2Se_3$ growth, and thus may offer advantages over sapphire as a substrate. InP(111) has been used before and reported by some to produce good material properties,[28,30,31] although with twin domains. There are a few reports of full twin suppression of $Bi_2Se_3$ by growing on 3.5° off-cut InP(111)A,[19] rough InP(111)A,[19] InP(115)[32], and rough InP(111)B[33] substrates. In the first two studies mentioned here, there is little information reported on the details of the structural analysis of the quality of the layers. In the case of InP(115),[32] a giant corrugation of the layers resulted in low structural quality and high strain which even modified the band structure of $Bi_2Se_3$. In the study of rough InP(111)B,[33] in which a detailed structural analysis has been done, a lower quality of $Bi_2Se_3$ is obtained compared to the previously reported high quality, but twinned $Bi_2Se_3$ on smooth InP(111)B,[31] as indicated by the broadening of the full width at half maximum (FWHM) of the rocking curve on rough substrates. Therefore, there is a critical need to improve the quality of these twin suppressed layers.

Although structurally InP(111)B is almost ideal for $Bi_2Se_3$ growth, the reactivity of the InP surface makes its preparation for growth more complex. As a method of compensating for the reactivity of Si(111) substrates, an $In_2Se_3$[25] buffer was used which improved the quality of the layer but did not help with twin suppression when grown on flat non-vicinal substrates. $In_2Se_3$ grown on Si(111) vicinal substrates[25] suppressed the twinning but gave rise to a corrugated surface. Therefore, the details of how we prepare the InP substrate before the growth of any layer are of utmost importance, and the influence of the substrate on the growth layer can be critical.

In this study we have investigated a novel method developed by us for the preparation of the non-vicinal InP(111)B substrate surface, and explored its properties with regard to formation of lower defect densities in the TI materials, in particular of twin domains. As the substrate preparation technique, we developed a new method that results in the formation of 2-dimentional (2D) van der-Waals $In_2Se_3$ ultra-thin layers via Se passivation during the oxide desorption process of the non-vicinal InP (111) substrate, in a molecular beam epitaxy (MBE) environment. The $In_2Se_3$ is formed without using an indium (In) source. With this new technique, we transformed the first few layers of the InP substrate into a van der-Waals $In_2Se_3$ layer which is structurally and chemically compatible with $Bi_2Se_3$. Due to this formation process, the resulting layers are fully twin suppressed.

It was previously shown that growing single phase $In_2Se_3$ is very difficult since $In_2Se_3$ has many different phases that can be formed readily[34]. Even when single phase $In_2Se_3$ was achieved, the layers were twinned.[25,35] Conventional MBE growth of $In_2Se_3$ would not produce un-twinned layers because there is freedom for the arriving indium atoms to move on the surface and form equally energetically-favorable twin domains. One way to obtain twin suppressed $In_2Se_3$ is to restrict the degrees of freedom of the indium atoms so that only one of the twin domains is favorable. Our approach of forming the $In_2Se_3$ layer by an exchange of InP(111) P atoms with Se atoms achieves this restriction, and leads to un-twinned $In_2Se_3$ layers, resulting in an ideal virtual substrate for the growth of $Bi_2Se_3$ and $Sb_2Te_3$. In this paper we present our surface modification technique and investigate the effect of this layer on the quality of epitaxial $Bi_2Se_3$ and $Sb_2Te_3$ grown on this virtual substrate. We present the growth details and the quality of the resulting $In_2Se_3$ and explore the crystal quality and twin-free nature of $Bi_2Se_3$ and $Sb_2Te_3$ layers grown on $In_2Se_3$/InP (111) virtual substrates by MBE.

**Results and Discussion**

**Formation of $In_2Se_3$ via selenium passivation of InP(111)B.** The key to the nucleation of a twin free $In_2Se_3$ layer on the InP substrate is the control of the nucleation process such that it avoids the coalescence of multiple nucleation sites into one continuous layer. Thus, the standard approach of deposition of In and Se atoms on the oxide free InP surface does not work. The approach taken here is to use the In from the InP(111) surface. In this process, the In atoms act as fixed "anchors" that ensure the layer is all equally aligned. The following steps were used to form the $In_2Se_3$ layer. First, the substrate temperature ($T_{sub}$) was raised to 200 °C and held for 5 minutes to stabilize the temperature; at this point the RHEED pattern was completely spotty (1x1) as shown in Fig. 1a. This is typical for an oxidized rough surface. A selenium overpressure of $1x10^{-5}$ Torr was provided at a substrate temperature of 300 °C and above, to prevent the out diffusion of phosphorus (P) atoms. Then, the temperature was raised gradually up to 495 °C at an average rate of 50°C per minute without annealing at any temperature; at this point the RHEED pattern had changed to a mix of streaks and spots (Fig. 1b), suggesting the onset of deoxidation. The substrate was annealed for 5 minutes at this temperature while observing the RHEED, which remained unchanged. After that, the temperature was increased to 505 °C and annealed for another 5 minutes. At this point the RHEED pattern became completely streaky (1x1) (Fig. 1c) indicating full oxide removal of the surface. As soon as the RHEED pattern changed to a fully streaky pattern, the power to the substrate heater was shutdown allowing a rapid cool down to 190 °C. After the oxide removal, the streaky (1x1) RHEED pattern remained unchanged. This pattern corresponds to the surface reconstruction of the newly formed $In_2Se_3$ layer during the oxide desorption process, without using an external indium source.

Once the temperature reached 190 °C, in order to deposit an epitaxial layer on the $In_2Se_3$ surface, either the Se flux was adjusted to the required amount for Bi$_2$Se$_3$ growth, or the Se shutter was closed for Sb$_2$Te$_3$ growth, and the substrate temperature was raised to the required growth temperature. The temperature of the substrate was held at the growth temperature for 10 minutes before starting the growth to ensure a clean surface with no excess Se on the surface.

**Structural properties of In$_2$Se$_3$/InP(111)B virtual substrate.** Fig. 1d shows an atomic force microscopy (AFM) image of a smooth, atomically flat In$_2$Se$_3$ layer formed during Se passivation of the InP(111)B substrate without an indium source. This layer has a roughness of Rq=0.4 nm. The XRD 2θ-ω scan of the In$_2$Se$_3$ layer which is shown in Fig. 3a, has well defined peaks from multiple reflections of (003) plane. The position of the peaks indicates that the sample is 3R In$_2$Se$_3$, a rhombohedral crystal structure; the data suggests it can be either α-In$_2$Se$_3$ ($R3m$) or β-In$_2$Se$_3$ ($R\bar{3}m$)[36,37]. The high crystallinity of the In$_2$Se$_3$ layer is evidenced by the thickness fringes of the (006) reflection from which the thickness was calculated. The inset in Fig. 3a is an X-ray Reflectivity (XRR) scan of the In$_2$Se$_3$ layer presenting pronounced oscillations; the thickness calculated using these oscillations is comparable to the value obtained using the thickness fringes in the XRD 2θ-ω scan (thickness of ~6nm). The full width at half maximum (FWHM) of the high-resolution X-ray diffraction (HRXRD) rocking curve is 0.06° (see Supplemental Information Fig. 1a) which is a good number for an ultra-thin layer. Fig. 3b shows the XRD ϕ scan of the (015) plane of two different In$_2$Se$_3$ samples with the characteristic three peaks, one peak at every 120°, indicating that the layer consists of a single domain. The FWHM of the XRD ϕ scan is 0.01° which confirms that the hexagonal lattice of In$_2$Se$_3$ is aligned parallel to that of the InP(111) within a twist angle as small as 0.01° (see Supplemental Information Fig. 2b). Fig. 3b shows that the three peaks do not always occur at the same in-plane angle as the (200) plane reflections of the substrate; rather in some cases the other twin orientation with respect to the substrate is preferred[28]. In both cases, the In$_2$Se$_3$ layer is un-twinned.

**Structural properties of the grown Bi$_2$Se$_3$, and Sb$_2$Te$_3$ epitaxial layers.** Bi$_2$Se$_3$ and Sb$_2$Te$_3$ layers were grown on these In$_2$Se$_3$/InP(111)B virtual substrate by MBE. The layers were grown at different growth temperatures ($T_g$ = 245 - 295°C) to establish its optimum value. An AFM image of an 18 nm thick Bi$_2$Se$_3$ film (sample S3) grown on the virtual substrate is shown in Fig. 2d. The surface is composed of large terraces with a very small roughness of $R_q$=0.5 nm. Fig. 4a shows XRD 2θ-ω scans of a series of Bi$_2$Se$_3$ films with different thicknesses grown at different $T_g$. All the samples show peaks due to multiple reflections of (003) plane from Bi$_2$Se$_3$ layers with pronounced thickness fringes in both (003) and (006) reflections indicating high material and interface quality. As Bi$_2$Se$_3$ is closely lattice matched to In$_2$Se$_3$ (both α and β phases) the peaks due to reflections from the underlying In$_2$Se$_3$ layer cannot be distinguished. However, the peak broadening around the base of the (0015) and (0018) reflections gives evidence of the existence of this layer. Fig. 4b shows XRD ϕ scans of the (015) plane of the three samples shown in Fig. 4a. Full twin suppression was achieved for the sample S3 with the highest growth temperature (295 °C) while nearly full twin suppression (~0.5% twinning) was achieved for all the other samples grown throughout this temperature range. Again, the three peaks from the (015) plane do not always occur at the same in-plane angle as the (200) plane reflections of the substrate; in some cases, the other twin orientation with respect to the substrate is preferred, as observed for the $In_2Se_3$ layer. However, in either case, a single crystal layer with full twin suppression was achieved. For all the samples, the FWHM of the HRXRD rocking curve of the $Bi_2Se_3$ films is ~0.06° (see Supplemental Information Fig. 1b), which is 5x less than the fully twin suppressed $Bi_2Se_3$ grown on roughened InP(111)B substrates reported in the literature[33]. The layer-twist of the samples is in the range of 0.2° – 0.5° (determined by the FWHM of the peak of XRD $\phi$ scan, see Supplemental Information Fig. 3b) is comparable to the values found in the literature for the twin suppressed samples grown on roughened InP(111)B substrate.[33] We conclude that the quality of the $Bi_2Se_3$ is improved significantly when grown on the smooth $In_2Se_3$ surfaces formed by our procedure rather than on the rough InP surfaces.

In comparison to $Bi_2Se_3$, $Sb_2Te_3$ layers grown on these virtual substrates, exhibit a larger degree of roughness, as is typical for $Sb_2Te_3$[38] Fig. 2e and 2f show AFM images of $Sb_2Te_3$ (sample S4 and S6) with a roughness of 2.2 nm and 3.8 nm respectively, When $Sb_2Te_3$ was grown on a $Bi_2Se_3$ layer grown previously on the $In_2Se_3$ virtual substrate, the surfaces became even rougher than the ones directly grown on the $In_2Se_3$ (Fig. 2f). XRD $2\theta$-$\omega$ scans of $Sb_2Te_3$ on $In_2Se_3$ layer and on the $Bi_2Se_3/In_2Se_3$ heterostructure are shown in Fig. 5a. Since the $In_2Se_3$ is not lattice matched to $Sb_2Te_3$, $2\theta$-$\omega$ scans of S4 and S5 samples show two distinct sets of multiple reflections of the (003) plane; one set is due to the $Sb_2Te_3$ film, and the other set is due to the underlying $In_2Se_3$ layer as marked in the Figure. XRD $2\theta$-$\omega$ scans of S6 shows two distinct oscillations due to $Bi_2Se_3$ and $Sb_2Te_3$ layers, but the reflections due to $In_2Se_3$ are overlapping with the reflections of the $Bi_2Se_3$. Thickness fringes of the $2\theta$-$\omega$ scans due to these thin $Sb_2Te_3$ films are not observed possibly due to high interface roughness. Fig. 5b shows XRR measurements of samples S4 and S5 which consist of oscillations with two distinct periods; oscillations with the short period are due to $Sb_2Te_3$ and the ones with the long period are due to $In_2Se_3$ from which we calculated the thickness of the two layers. Fig. 6 shows $\phi$ scans of (015) plane of these samples. Full twin suppression was achieved for $Sb_2Te_3$ grown on $In_2Se_3$ layers (sample S4) whereas $Sb_2Te_3$ grown on the $Bi_2Se_3/In_2Se_3$ heterostructure had a small degree (~1%) of twinning. The diminished quality of the $Sb_2Te_3$ grown on $Bi_2Se_3$ may be due to a non-optimal $T_g$ that was used as a compromise for the structure, i.e., the growth temperature used is not the best for either layer. Optimization of the growth conditions may further reduce the twinning of these complex structures, which is desirable for the growth of $Sb_2Te_3/Bi_2Se_3$ superlattice structures with full twin suppression conditions. These superlattices have been shown to produce lower carrier density materials due to superlattice gap enhancements.[39]

**Transport properties of $Bi_2Se_3$ and $Sb_2Te_3$.** Surface preparation and substrate/epilayer interface are known to affect the transport properties of the TI layers. Fig. 7a presents the thickness dependence of the 2D carrier density of $Bi_2Se_3$ and $Sb_2Te_3$ on $In_2Se_3$ virtual substrate. The data show that the variation in 2D carrier density is small, and there is no correlation between the layer thickness and the 2D carrier density of these materials. The average 2D carrier density of $Bi_2Se_3$ is ~ $2\times10^{13}$ cm$^{-2}$ (n-type) whereas for $Sb_2Te_3$ it is ~$1\times10^{14}$ cm$^{-2}$ (p-type), an order of magnitude higher. The average 2D carrier density of $Bi_2Se_3$ is ~2.5 times lower than the twinned $Bi_2Se_3$ grown directly on variety of substrates, which are previously reported[40,41,42,43]. For comparison, we have shown the data for $Bi_2Se_3$ grown on $Al_2O_3$ in our MBE chamber which had only 10% of twinning[44] as a red triangle on the plot. The previously reported lowest 2D carrier density for the fully twin suppressed $Bi_2Se_3$ grown on roughened InP(111)B[45] is $0.9\times10^{13}$ cm$^{-2}$, comparable to the lowest value we measured for twin suppressed $Bi_2Se_3$, which is $1\times10^{13}$ cm$^{-2}$. Therefore, we conclude that the electrical properties of $Bi_2Se_3$ samples grown on the $In_2Se_3$/InP(111)B virtual substrates are of comparable quality to the ones reported in the literature with low twinning or fully twin suppressed samples. On the other hand, 2D carrier density of our untwinned $Sb_2Te_3$ does not decrease compared to the twinned samples (red circle on the plot shows the 2D carrier density of $Sb_2Te_3$ grown on $Al_2O_3$ in the same MBE system), probably because the contribution from the other defects outweighs the ones from the suppressed twinning. The observation that the 2D carrier density is independent of the layer thickness suggests that the dominant contribution of bulk carriers in our samples originates from the interface thus the defect density in the bulk is constant. Similar observations were previously reported for these material grown on $Al_2O_3$[26]. Fig. 7b shows the thickness dependence of the mobility of $Bi_2Se_3$ and $Sb_2Te_3$ layers. Carrier mobility of $Sb_2Te_3$ exhibits a steep decrease below ~40 nm whereas $Bi_2Se_3$ has a slow and gradual decrease. This indicates that the $Sb_2Te_3$/$In_2Se_3$ interface has a larger density of scattering centers compared to the $Bi_2Se_3/In_2Se_3$ interface. This is likely due to the lattice mismatch at the $Sb_2Te_3/In_2Se_3$ interface, which may lead to more scattering centers. It may also be possible that there is intermixing or reacting of In and Se atoms with the $Sb_2Te_3$ layer at the interface. For thicker layers (layers thicker than 40 nm), carrier mobility of $Sb_2Te_3$ grown on $Al_2O_3$ falls along the same line as the twin suppressed samples, as shown by the red circle in Fig. 7b. Further, the mobility of the 10% twinned $Bi_2Se_3$ on $Al_2O_3$ follows the mobility trend of the fully twin suppressed $Bi_2Se_3/In_2Se_3$ layers presented here. Since most carriers originate at the interface, it is reasonable to expect that the use of a thicker $In_2Se_3$ buffer layer would result in lower carrier densities in the TI layers.

**Conclusions**

In this study, we developed a novel selenium passivation technique to convert the 3D non-vicinal InP(111)B substrate surface into an $In_2Se_3$ 2D van der Waals virtual substrate without using an indium cell. Smooth ultra-thin (~6 nm) $In_2Se_3$ layers were grown reproducibly using this new technique. The layers have good crystalline quality and exhibit no twinning. We have also shown that full twin suppression of $Bi_2Se_3$ and $Sb_2Te_3$ layers was achieved by growth on non-vicinal InP(111)B via the use of the $In_2Se_3$/InP(111) B virtual substrate. These $Bi_2Se_3$ and $Sb_2Te_3$ layers have greatly improved crystalline quality over the twin suppressed ones grown on roughened InP(111)B substrates. Furthermore, the electrical properties of these layers are comparable to those reported for other twin suppressed layers. We propose that the absence of twins is achieved by suppressing In-atom mobility on the surface, which allows all the $In_2Se_3$ nucleation sites to be aligned to the InP(111) substrate, thus eliminating the possibility of twinned domains. Twin suppressed TI layers have potential advantages for novel device applications that rely on selective interactions between polarized light and the spin helicity properties of the topological surface states. For example, it has been shown that twin suppression increases the helicity-dependent photo response which is predicted as a potential application in chip-scale polarimeters[14]. Hence, high quality TIs with full twin suppression greatly broadens possible device applications in the terahertz regime.

**Methods**

All samples were grown on non-vicinal Fe doped InP(111)B $\pm$ 0.5° substrates, which have a phosphorous terminated surface. A Riber 2300P system with a base pressure of $5\times10^{11}$ Torr was used, equipped with *in-situ* reflection high-energy electron diffraction (RHEED) to monitor the growth of the material in real time. High-purity (99.9999%) bismuth (Bi) and antimony (Sb) fluxes were provided by a RIBER dual zone effusion cell while a RIBER valved cracker cell was used for selenium (Se) and a single zone Knudsen cell for tellurium (Te). Beam equivalent pressure (BEP) was measured by an ultra-high vacuum (UHV) nude ion gauge placed in the path of the fluxes. All temperatures reported here are measured using thermocouple.

**Growth of $Bi_2Se_3$ and $Sb_2Te_3$ on $In_2Se_3$/InP(111)B.** A series of $Bi_2Se_3$ and $Sb_2Te_3$ samples were grown on $In_2Se_3$ layer to optimize and investigate the structural and electrical properties of these materials. The $Bi_2Se_3$ growth temperature was varied in the range of $T_{sub}$ = 245-295 °C while a fixed growth rate of ~0.5 nm/minute and a BEP ratio of Se to Bi ~100:1 were maintained. A 1x1 RHEED pattern was evident after $Bi_2Se_3$ growth with an enhanced streak intensity relative to the $In_2Se_3$ layer, as shown in Fig. 2a. The $Sb_2Te_3$ growth temperature was varied in the range of $T_{sub}$ = 235-260 °C with a growth rate ~0.25 nm/minute and a BEP ratio of Te to Sb ~15:1. The 1x1 RHEED pattern after $Sb_2Te_3$ growth is shown in Fig. 2b. A layer of $Sb_2Te_3$ was also grown on a $Bi_2Se_3$(18 nm) layer grown on the $In_2Se_3$ surface at a growth temperature of $T_{sub}$ = 250 °C with a $Bi_2Se_3$ and $Sb_2Te_3$. BEP ratios of Se to Bi ~100:1 and Te to Sb ~ 15:1 were used. After growth of $Sb_2Te_3$/$Bi_2Se_3$, a 1x1 RHEED pattern was observed, as shown in Fig. 2c. The post growth RHEED intensity in both cases of the $Sb_2Te_3$ growth decreased compared to the RHEED intensity of the $In_2Se_3$ layer.

**Structural and Electrical Characterization.** A Bruker D8 Discover diffractometer with a da Vinci configuration and a conditioned Cu K$\alpha_1$(1.5418 Å) source was used for The XRD measurements. A Bruker Dimension FastScan AFM with a FastScan-A silicon probe was used to capture the Atomic Force Microscopy (AFM) images. Surface roughness (root mean square roughness, $R_q$) of the samples was determined using 4µm$^2$ AFM scans. Transport properties were measured in van der Pauw geometry using four-wire measurements in a closed-cycle helium cryostat at 10K. Square pieces of ~5mm x 5mm were used for the measurement and the electrical contacts were made using pure indium.

**Data availability**

All data are available from the corresponding author upon reasonable request.

**References**

1. Xia, Y. *et al.* Observation of a large-gap topological-insulator class with a single Dirac cone on the surface. *Nat. Phys.* **5**, 398–402 (2009).

2. Zhang, H. *et al.* Topological insulators in Bi2Se3, Bi2Te3 and Sb2Te3 with a single Dirac cone on the surface. *Nat. Phys.* **5**, 438–442 (2009).

3. Hasan, M. Z. & Kane, C. L. Colloquium: Topological insulators. *Rev. Mod. Phys.* **82**, 3045–3067 (2010).

4. Qi, X.-L. & Zhang, S.-C. The quantum spin Hall effect and topological insulators. *Phys. Today* **63**, 33–38 (2010).

5. Guo, M. *et al.* Tuning thermoelectricity in a Bi2Se3 topological insulator via varied film thickness. *New J. Phys.* **18**, 015008 (2016).

6. He, Q. L., Hughes, T. L., Armitage, N. P., Tokura, Y. & Wang, K. L. Topological spintronics and magnetoelectronics. *Nat. Mater.* **21**, 15–23 (2022).

7. Wu, F., Zhang, R.-X. & Das Sarma, S. Three-dimensional topological twistronics. *Phys. Rev. Res.* **2**, 022010 (2020).

8. Zhou, C., Song, D., Jiang, Y. & Zhang, J. Modification of the Hybridization Gap by Twisted Stacking of Quintuple Layers in a Three-Dimensional Topological Insulator Thin Film. *Chin. Phys. Lett.* **38**, 057307 (2021).

9. Fu, L. & Kane, C. L. Superconducting Proximity Effect and Majorana Fermions at the Surface of a Topological Insulator. *Phys. Rev. Lett.* **100**, 096407 (2008).

10. Tian, W., Yu, W., Shi, J. & Wang, Y. The Property, Preparation and Application of Topological Insulators: A Review. *Materials* **10**, (2017).

11. Chang, C.-Z., Liu, C.-X. & MacDonald, A. H. Colloquium: Quantum anomalous Hall effect. *Rev. Mod. Phys.* **95**, 011002 (2023).

12. Yao, J. D., Shao, J. M., Li, S. W., Bao, D. H. & Yang, G. W. Polarization dependent photocurrent in the Bi2Te3 topological insulator film for multifunctional photodetection. *Sci. Rep.* **5**, 14184 (2015).

13. Tu, C.-M. *et al.* Helicity-dependent terahertz emission spectroscopy of topological insulator Sb2Te3 thin films. *Phys. Rev. B* **96**, 195407 (2017).

14. Connelly, B. C., de Coster, G. J. & Taylor, P. J. Helicity- and Azimuthal-Dependent Topological Photocurrents in Bi2Se3 using THz Spectroscopy. in *Conference on Lasers and Electro-Optics* STh2O.2 (Optica Publishing Group, 2022). doi:10.1364/CLEO_SI.2022.STh2O.2.

15. Hsieh, D. *et al.* A tunable topological insulator in the spin helical Dirac transport regime. *Nature* **460**, 1101–1105 (2009).

16. Chen, J. *et al.* Gate-Voltage Control of Chemical Potential and Weak Antilocalization in Bi2Se3. *Phys. Rev. Lett.* **105**, 176602 (2010).

17. Richardella, A. *et al.* Coherent heteroepitaxy of Bi2Se3 on GaAs (111)B. *Appl. Phys. Lett.* **97**, 262104 (2010).

18. Ginley, T. P., Wang, Y. & Law, S. Topological Insulator Film Growth by Molecular Beam Epitaxy: A Review. *Crystals* **6**, (2016).

19. Guo, X. *et al.* Single domain Bi2Se3 films grown on InP(111)A by molecular-beam epitaxy. *Appl. Phys. Lett.* **102**, 151604 (2013).

20. Koma, A. Van der Waals epitaxy for highly lattice-mismatched systems. *J. Cryst. Growth* **201–202**, 236–241 (1999).

21. He, L. *et al.* Epitaxial growth of Bi2Se3 topological insulator thin films on Si (111). *J. Appl. Phys.* **109**, 103702 (2011).

22. West, D., Sun, Y. Y., Wang, H., Bang, J. & Zhang, S. B. Native defects in second-generation topological insulators: Effect of spin-orbit interaction on Bi2Se3. *Phys. Rev. B* **86**, 121201 (2012).

23. Kampmeier, J. *et al.* Selective area growth of Bi2Te3 and Sb2Te3 topological insulator thin films. *J. Cryst. Growth* **443**, 38–42 (2016).

24. Bansal, N. *et al.* Epitaxial growth of topological insulator Bi2Se3 film on Si(111) with atomically sharp interface. *Thin Solid Films* **520**, 224–229 (2011).

25. Wang, Z. Y., Li, H. D., Guo, X., Ho, W. K. & Xie, M. H. Growth characteristics of topological insulator Bi2Se3 films on different substrates. *J. Cryst. Growth* **334**, 96–102 (2011).

26. Koirala, N. *et al.* Record Surface State Mobility and Quantum Hall Effect in Topological Insulator Thin Films via Interface Engineering. *Nano Lett.* **15**, 8245–8249 (2015).

27. Wang, Y., Ginley, T. P. & Law, S. Growth of high-quality Bi2Se3 topological insulators using (Bi1-xInx)2Se3 buffer layers. *J. Vac. Sci. Technol. B* **36**, 02D101 (2018).

28. Richardella, A., Kandala, A., Lee, J. S. & Samarth, N. Characterizing the structure of topological insulator thin films. *APL Mater.* **3**, 083303 (2015).

29. Tarakina, N. V. *et al.* Comparative Study of the Microstructure of Bi2Se3 Thin Films Grown on Si(111) and InP(111) Substrates. *Cryst. Growth Des.* **12**, 1913–1918 (2012).

30. Chen, Z. *et al.* Molecular Beam Epitaxial Growth and Properties of Bi2Se3 Topological Insulator Layers on Different Substrate Surfaces. *J. Electron. Mater.* **43**, 909–913 (2014).

31. Schreyeck, S. *et al.* Molecular beam epitaxy of high structural quality Bi2Se3 on lattice matched InP(111) substrates. *Appl. Phys. Lett.* **102**, 041914 (2013).

32. Takagaki, Y., Jenichen, B. & Tominaga, J. Giant corrugations in Bi2Se3 layers grown on high-index InP substrates. *Phys. Rev. B* **87**, 245302 (2013).

33. Tarakina, N. V. *et al.* Suppressing Twin Formation in Bi2Se3 Thin Films. *Adv. Mater. Interfaces* **1**, 1400134 (2014).

34. Liu, L. *et al.* Atomically Resolving Polymorphs and Crystal Structures of In2Se3. *Chem. Mater.* **31**, 10143–10149 (2019).

35. Claro, M. S., Grzonka, J., Nicoara, N., Ferreira, P. J. & Sadewasser, S. Wafer-Scale Fabrication of 2D β-In2Se3 Photodetectors. *Adv. Opt. Mater.* **9**, 2001034 (2021).

36. Tang, C., Sato, Y., Watanabe, K., Tanabe, T. & Oyama, Y. Selective crystal growth of indium selenide compounds from saturated solutions grown in a selenium vapor. *Results Mater.* **13**, 100253 (2022).

37. Küpers, M. *et al.* Controlled Crystal Growth of Indium Selenide, In2Se3, and the Crystal Structures of α-In2Se3. *Inorg. Chem.* **57**, 11775–11781 (2018).

38. Zeng, Z. *et al.* Molecular beam epitaxial growth of Bi2Te3 and Sb2Te3 topological insulators on GaAs (111) substrates: a potential route to fabricate topological insulator p-n junction. *AIP Adv.* **3**, 072112 (2013).

39. Levy, I. *et al.* Designer Topological Insulator with Enhanced Gap and Suppressed Bulk Conduction in Bi2Se3/Sb2Te3 Ultrashort-Period Superlattices. *Nano Lett.* **20**, 3420–3426 (2020).

40. Li, H. D. *et al.* The van der Waals epitaxy of Bi2Se3 on the vicinal Si(111) surface: an approach for preparing high-quality thin films of a topological insulator. *New J. Phys.* **12**, 103038 (2010).

41. Wang, X. *et al.* Transport properties of topological insulator Bi2Se3 thin films in tilted magnetic fields. *Phys. E Low-Dimens. Syst. Nanostructures* **46**, 236–240 (2012).

42. Glinka, Y. D., Babakiray, S. & Lederman, D. Plasmon-enhanced electron-phonon coupling in Dirac surface states of the thin-film topological insulator Bi2Se3. *J. Appl. Phys.* **118**, 135713 (2015).

43. Kim, N. *et al.* Persistent Topological Surface State at the Interface of Bi2Se3 Film Grown on Patterned Graphene. *ACS Nano* **8**, 1154–1160 (2014).

44. Levy, I., Garcia, T. A., Shafique, S. & Tamargo, M. C. Reduced twinning and surface roughness of Bi2Se3 and Bi2Te3 layers grown by molecular beam epitaxy on sapphire substrates. *J. Vac. Sci. Technol. B* **36**, 02D107 (2018).

45. 2D carrier density is calculated using the 3D carrier density of 100 nm thick Bi2Se3 in the reference Tarakina et al., 2014.

**Acknowledgements**

This work was supported by NSF grant numbers HRD-1547830 and HRD-2112550 (NSF CREST Center IDEALS). Partial support is also acknowledged from NSF grant number DMR-2011738 (PAQM). We would also like to acknowledge Dr. Thor Axtmann Garcia for helpful discussions, and the Nanofabrication Facility of CUNY Advanced Science Research Center (ASRC) for instrument use, scientific and technical assistance.

**Author contributions**

K.W. conceived and executed the research and MBE growth. K.W. and C.F. conducted the HR-XRD, XRR, AFM, and Hall transport measurements. K.W. and M.C.T. wrote and reviewed the manuscript. All authors contributed to interpretation of the data and discussions.

**Competing interests**

The authors declare no competing interests.

**Materials and Correspondence**

Correspondence and request for materials should be addressed to M.C.T

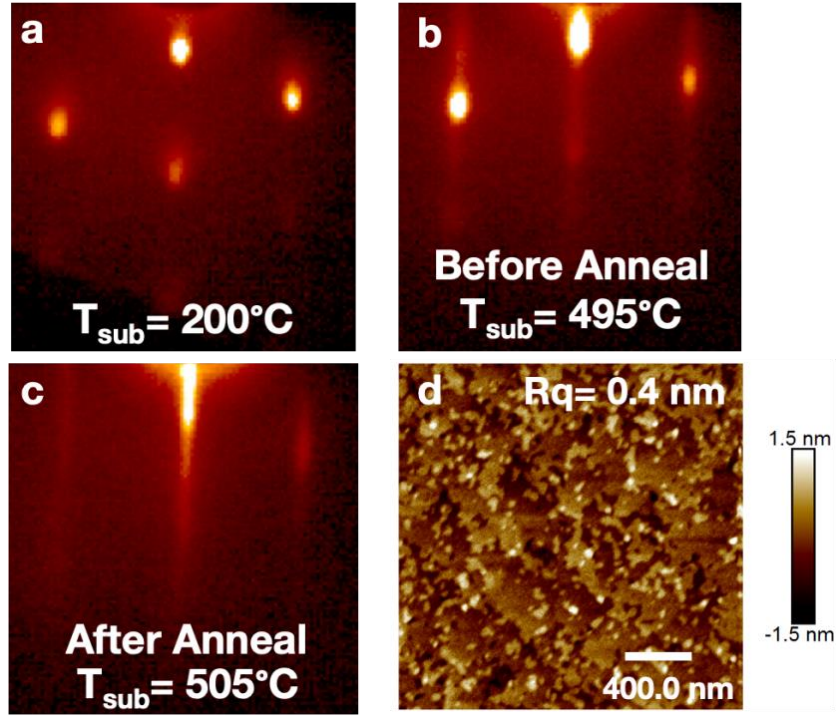

FIG. 1. Reflection high energy electron diffraction (RHEED) images of the InP(111)B surface (a) at $T_{sub}$=200 °C, (b) at $T_{sub}$=495 °C before anneal, (c) at $T_{sub}$=505 °C after 5 minutes anneal, i.e., the deoxidized surface with the newly formed $In_2Se_3$ layer via Se passivation of InP(111)B substrate. RHEED patterns are obtained along [0$\bar{1}$1] direction of the substrate. (d) Atomic force microscopy (AFM) image of the sample shown in (c) with smooth atomically flat $In_2Se_3$ layer which has a roughness of Rq=0.4 nm.

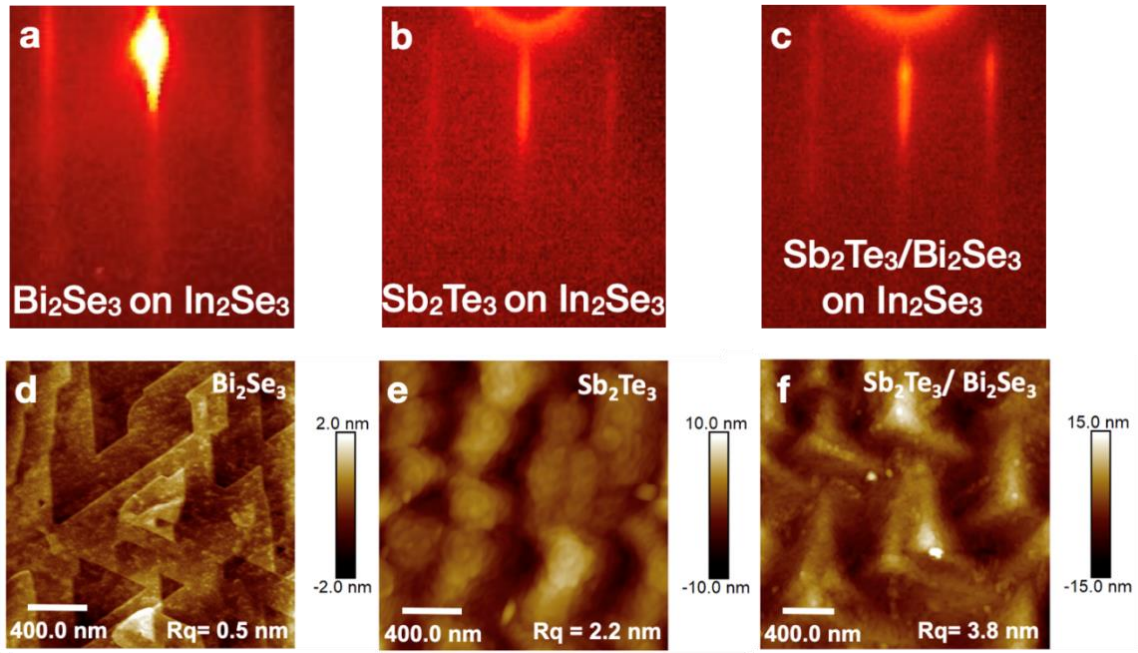

FIG. 2. Reflection high energy electron diffraction (RHEED) images of (a) after Bi$_2$Se$_3$ growth, (b) after Sb$_2$Te$_3$ growth, (c) after the growth of Sb$_2$Te$_3$/Bi$_2$Se$_3$ stack. RHEED patterns are obtained along [0$\bar{1}$1] direction of the substrate. (e) A 18 nm thick Bi$_2$Se$_3$ film (sample S3) with a roughness of Rq=0.5 nm. (f) A 29 nm thick Sb$_2$Te$_3$ film with a roughness of Rq=2.2 nm (sample S4), and (d) Sb$_2$Te$_3$/Bi$_2$Se$_3$ stack with a roughness of Rq=3.8 nm (sample S6).

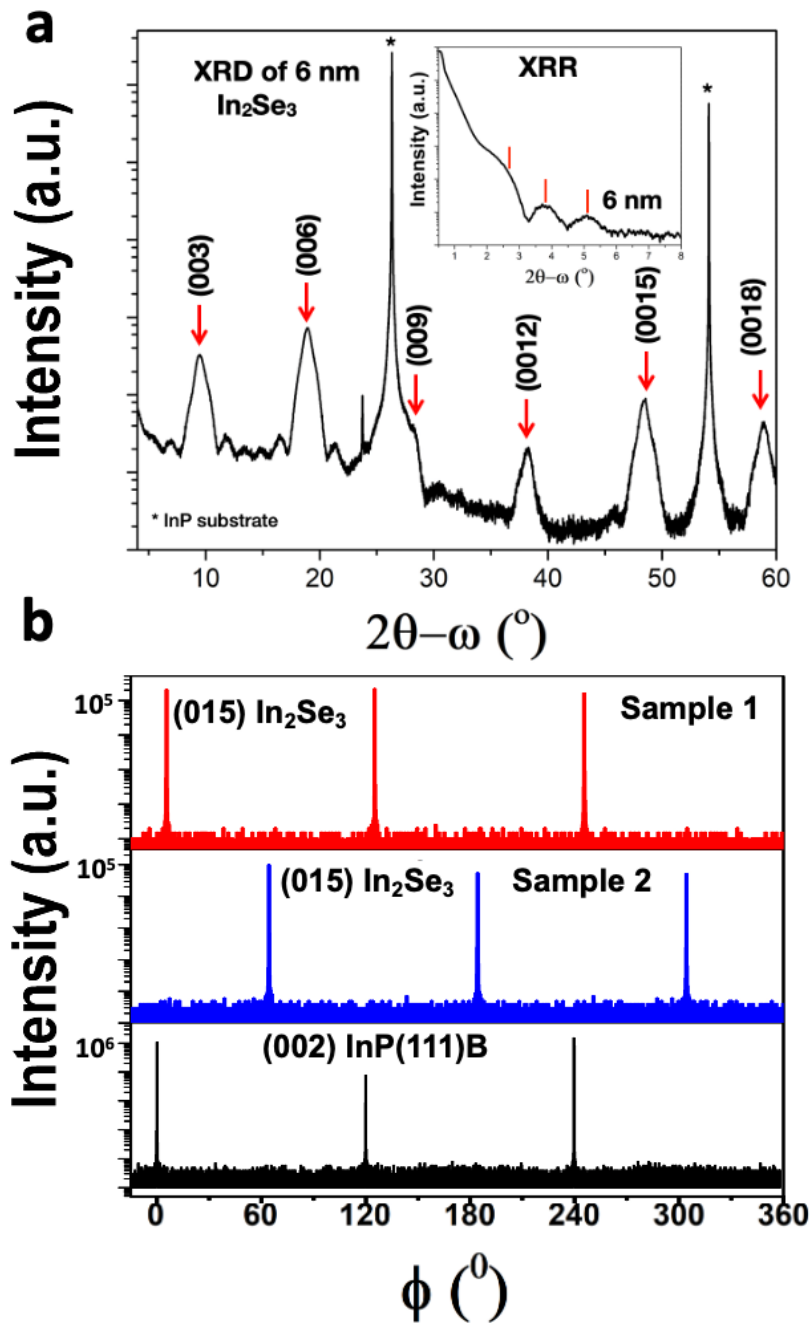

FIG. 3. X-ray diffraction (a) 2θ-ω scan of 6 nm thick In$_2$Se$_3$ layer; the inset shows X-ray reflectivity (XRR) of the In$_2$Se$_3$ layer with distinguishable oscillations. (b) φ scan of (015) plane of two different samples of In$_2$Se$_3$ and the (002) plane of InP(111)B substrate.

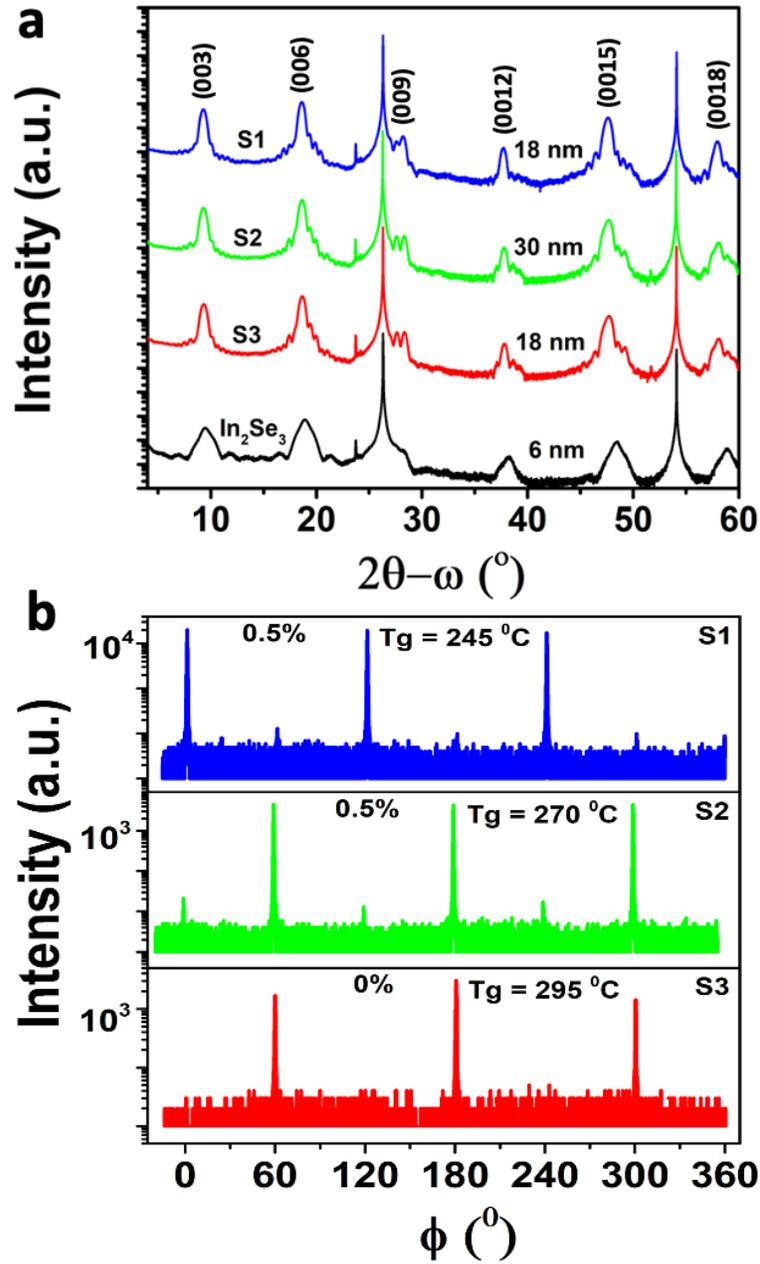

FIG. 4. X-ray diffraction (XRD) (a) 2θ-ω scans of Bi$_2$Se$_3$ films with different thicknesses grown at different substrate temperatures (T$_g$) on In$_2$Se$_3$ layer. (b) φ scan of (015) plane of the samples shown in (a) with the calculated twinning percentages.

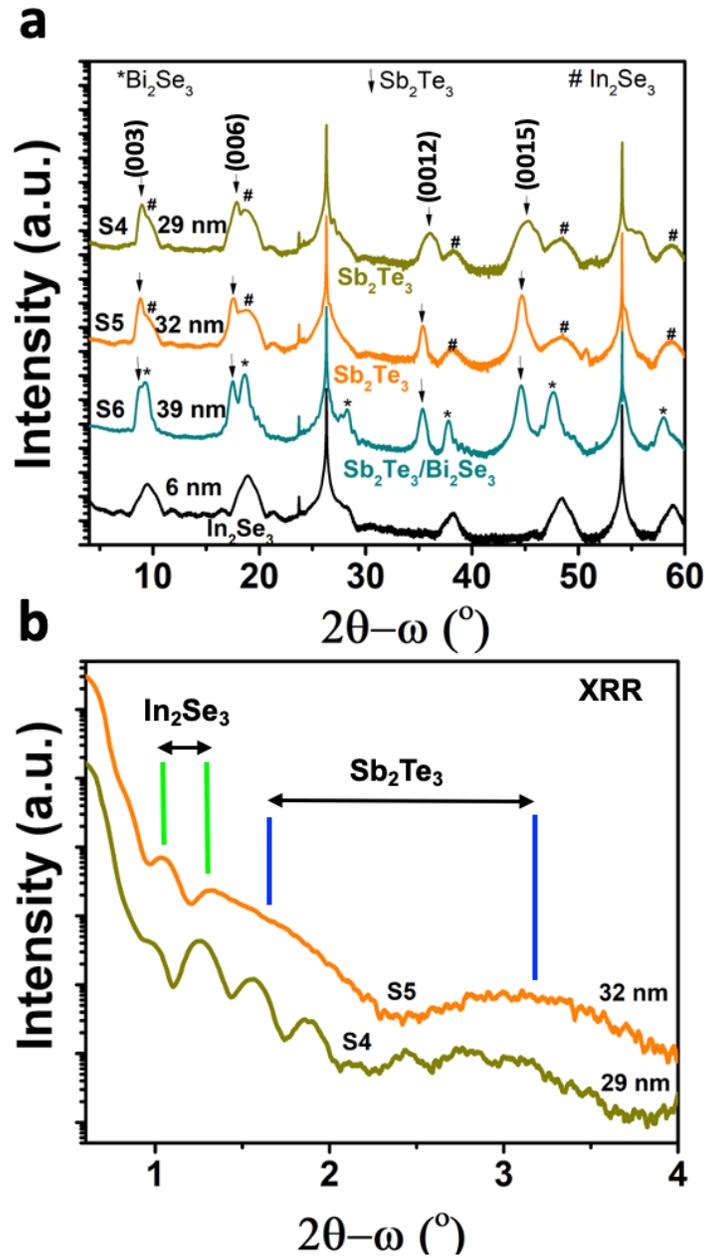

FIG. 5. X-ray diffraction (XRD) (a) 2θ-ω scans of $Sb_2Te_3$ films with different thicknesses grown at different substrate temperatures ($T_g$) on $In_2Se_3$ layer. (b) Sample X-ray reflectivity (XRR) scan of $Sb_2Te_3$ films showing two distinct oscillations corresponding to $In_2Se_3$ and the $Sb_2Te_3$ from which the layer thicknesses are calculated.

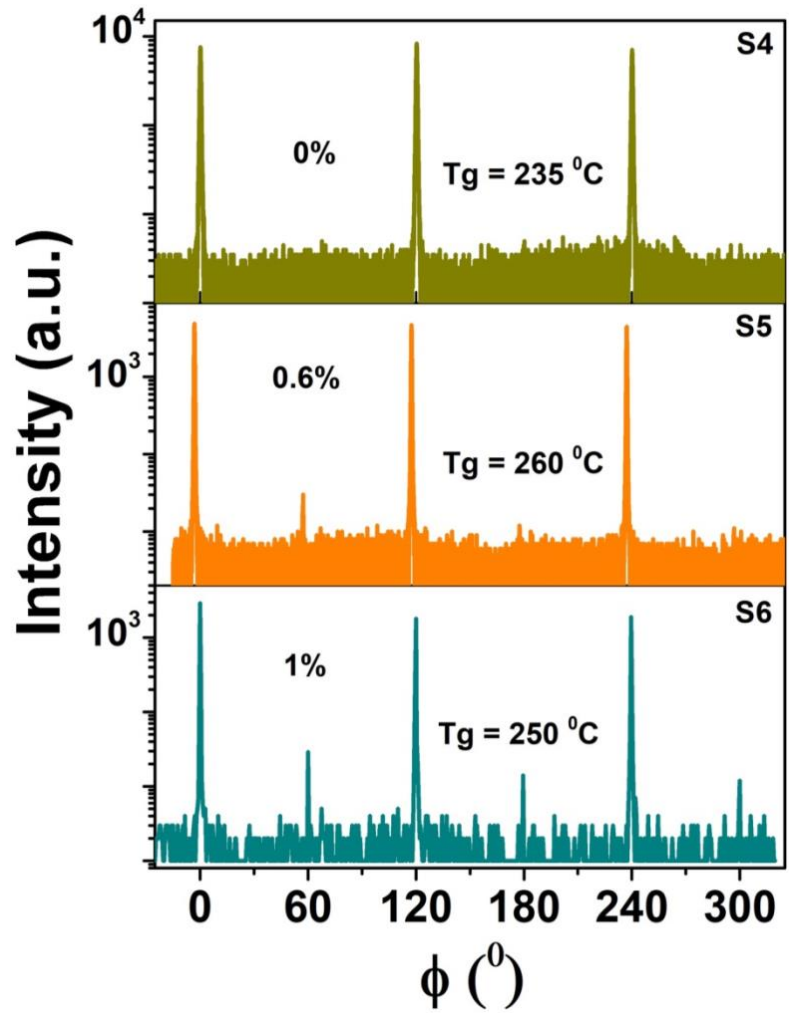

FIG. 6. ϕ scan of (015) plane of the $Sb_2Te_3$ and $Sb_2Te_3/Bi_2Se_3$ samples shown in Fig. 5a with the calculated twinning percentages. The dominant set of triplets shown here occur at the same in-plane angle as the (200) plane reflections of the substrate.

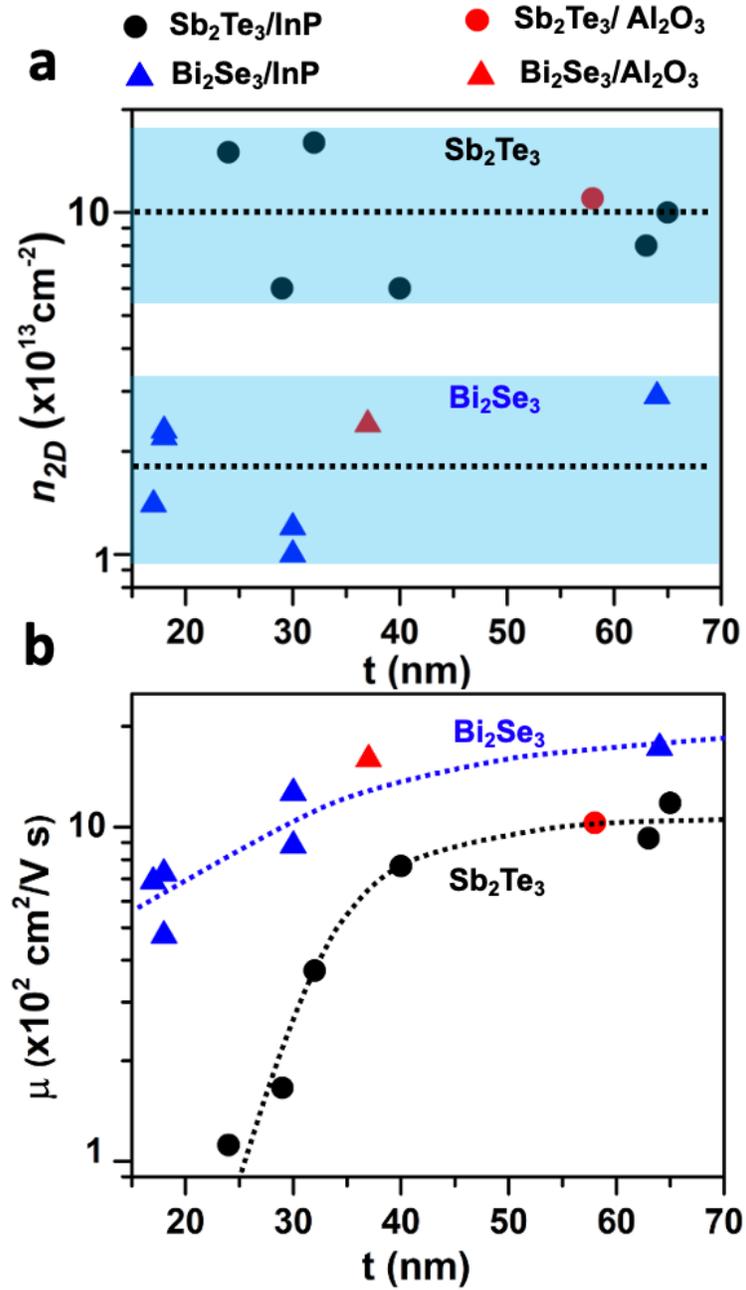

FIG. 7. (a) 2D carrier density and (b) Hall mobility of $Bi_2Se_3$ and $Sb_2Te_3$ on $In_2Se_3$/InP(111)B as a function of film thickness. Red circle and the triangle are representative values of films grown on $Al_2O_3$ for film thickness of 37 nm $Bi_2Se_3$ and 58 nm $Sb_2Te_3$ respectively. Measurements are taken at 10K.

**Supplementary Information**

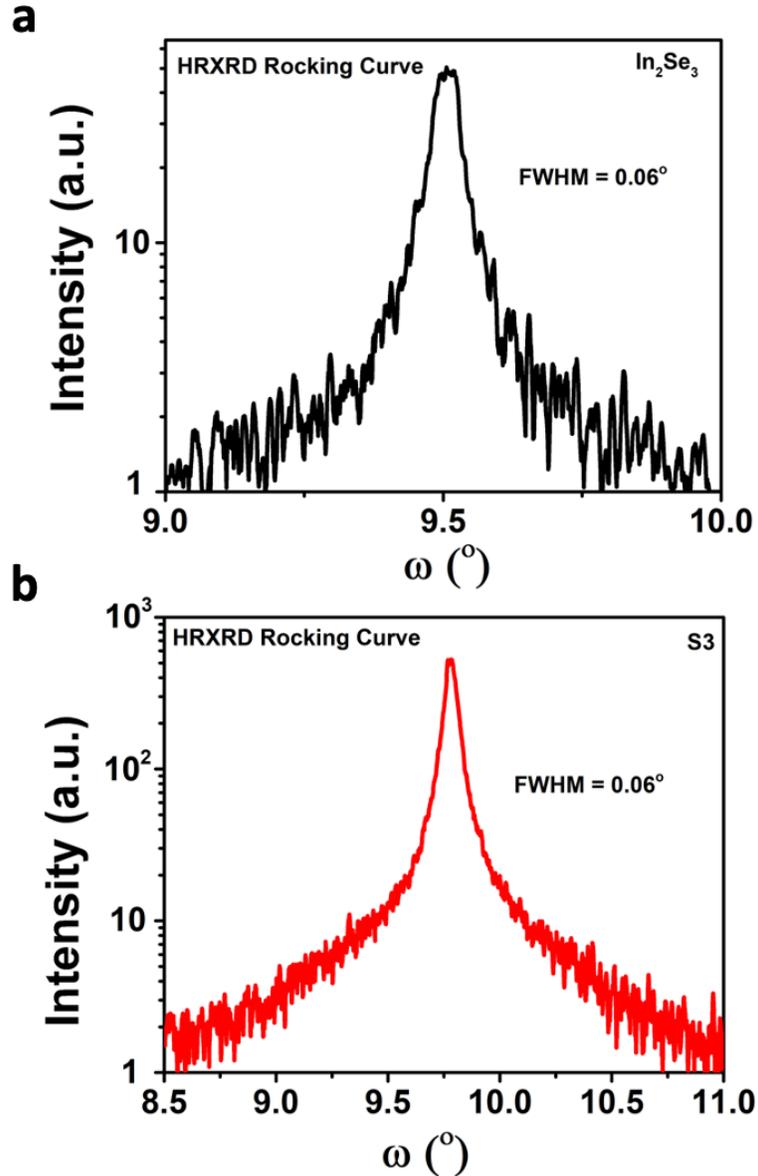

SUPPLEMENTARY FIG. 1. High resolution X-ray diffraction rocking curve (RC) of (006) plane of: (a) 6 nm thick $In_2Se_3$ layer on InP(111)B substrate (i.e., $In_2Se_3$ virtual substrate) with a full width at half maximum (FWHM) of 0.06° and (b) 18 nm thick $Bi_2Se_3$ layer (Sample 3) grown on an $In_2Se_3$ virtual substrate with a FWHM of 0.06°.

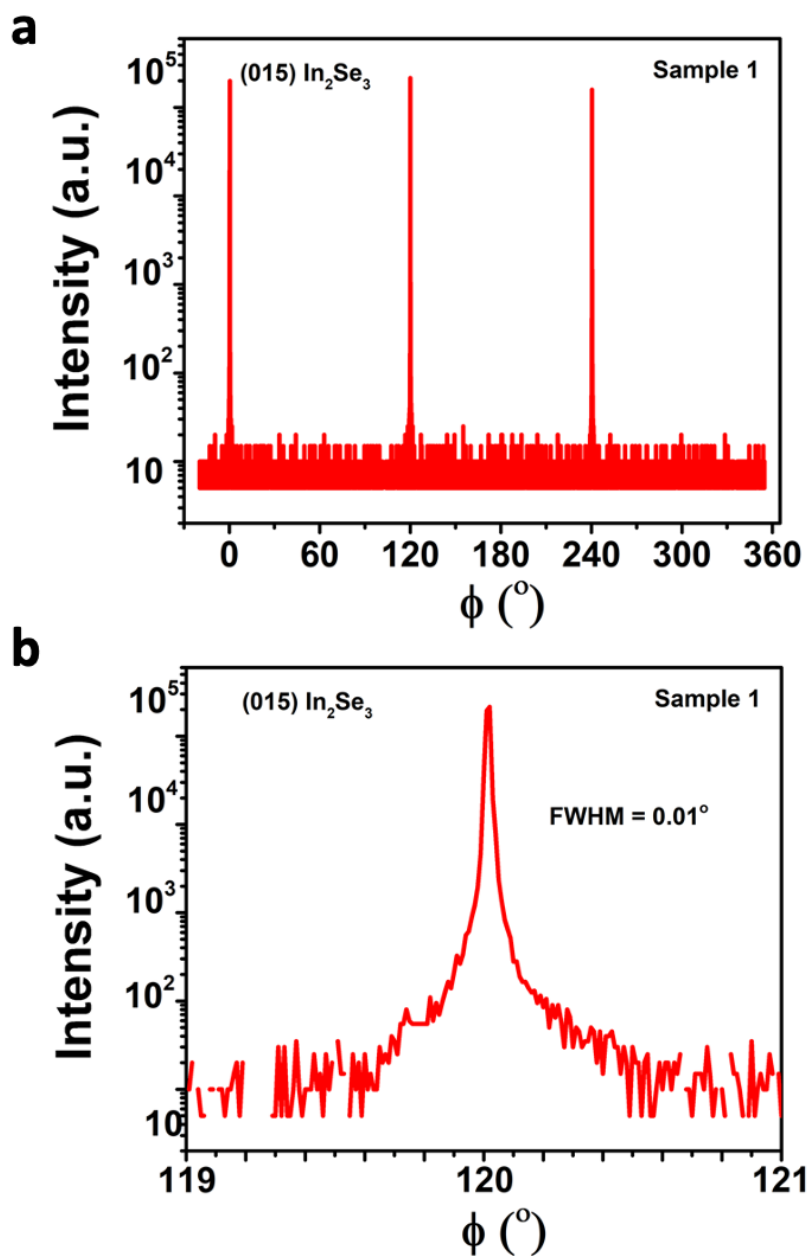

SUPPLEMENTARY FIG. 2. (a) $\phi$ scan of (015) plane of the un-twinned $In_2Se_3$ Sample 1. (b) Expanded view of the peak at 120° shown in (a) with a FWHM of 0.01°.

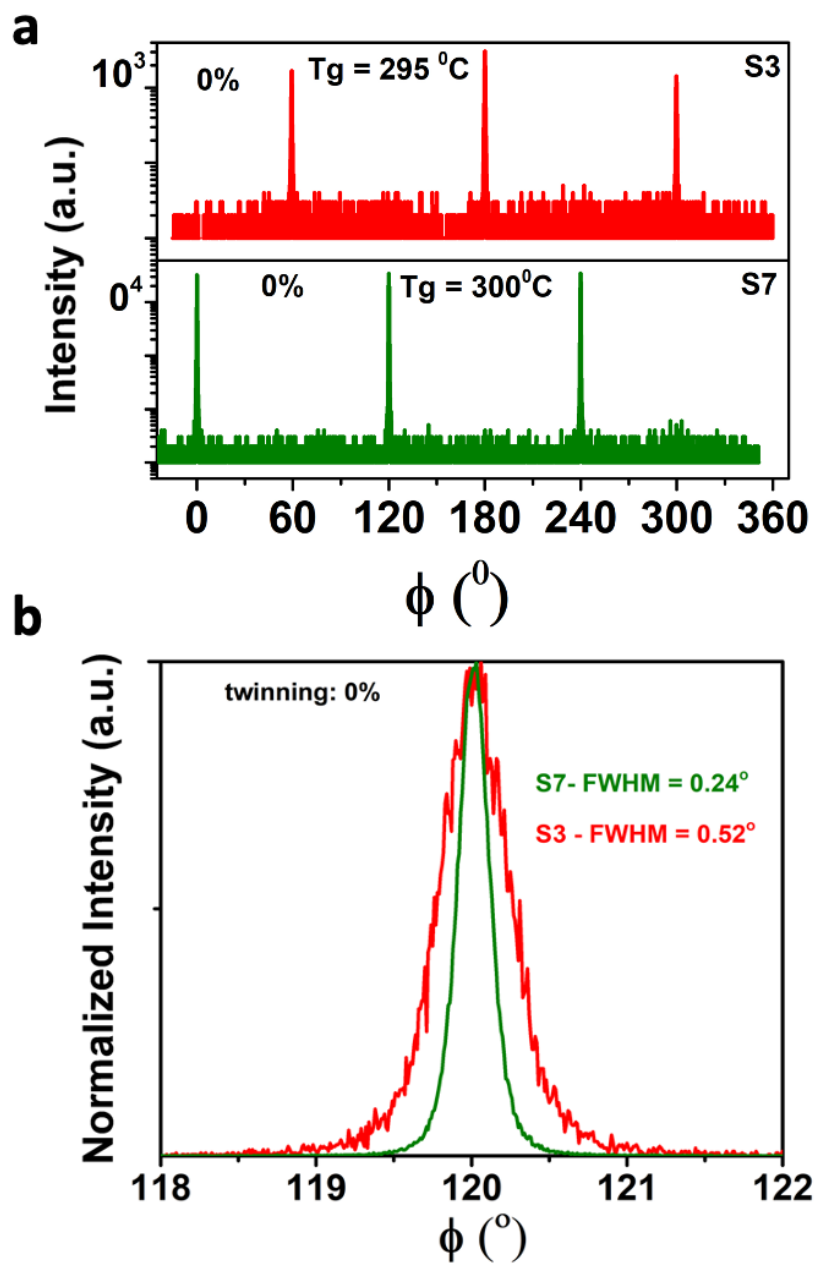

SUPPLEMENTARY FIG. 3. (a) φ scans of (015) plane of the fully twin suppressed Bi$_2$Se$_3$ Samples S3 (18 nm thick) and S7 (64 nm thick). (b) Expanded view of the middle peak of the triplet shown in (a). The middle peak of S3 shown in (b) is shifted -60° to align with the middle peak of S7 for the convenience of comparison.